\documentclass[letterpaper, 10 pt, conference]{ieeeconf} %
\IEEEoverridecommandlockouts                   %
\usepackage{cite}
\usepackage[para]{footmisc}
\usepackage{amsmath,amssymb,amsfonts}
\usepackage{algorithmic}
\usepackage{graphicx}
\graphicspath{{./img/}}
\usepackage{textcomp}
\usepackage[table,xcdraw]{xcolor}
\usepackage{diagbox}
\usepackage{colortbl}
\usepackage{multirow}
\usepackage{url}
\usepackage{nohyperref}
\usepackage[per-mode=symbol,group-separator={,},group-minimum-digits={3},binary-units]{siunitx}
\usepackage{subfig}
\usepackage[nohyperlinks,nolist]{acronym} 
\usepackage[]{algorithm2e}
\usepackage{tikz}
\usepackage{tabularx}
\newcolumntype{Y}{>{\centering\arraybackslash}X}
\newcolumntype{L}{>{\arraybackslash}X}
\newcolumntype{R}{>{\raggedleft\arraybackslash}X}
\newcolumntype{C}[1]{>{\centering\arraybackslash}p{#1}}
\usepackage{booktabs}
\usepackage[absolute,showboxes]{textpos}
\usetikzlibrary{patterns}
\usetikzlibrary{spy}
\usepackage{pgfplots}
\pgfplotsset{compat=newest}
\usepgfplotslibrary{units}
\usepgfplotslibrary{groupplots}
\makeatletter
\pgfplotsset{
    table/search path={data},
}
\pgfplotsset{
 unit code/.code 2 args=
   \begingroup
   \protected@edef\x{\endgroup\si{#2}}\x
}
\makeatother
\usetikzlibrary{matrix}
\usepackage[outline]{contour}
\usepackage{tcolorbox}
\tcbuselibrary{skins,breakable}
\usetikzlibrary{calc}

\definecolor{CoreGray}{HTML}{BFBFBF}
\definecolor{CoreBlack}{HTML}{333333}
\definecolor{CoreBlue}{HTML}{002E7D}
\definecolor{CoreGreen}{HTML}{6AAC8E}
\definecolor{CoreRed}{HTML}{C80000}
\definecolor{CoreYellow}{HTML}{E6AC00}
\definecolor{CoreWhite}{HTML}{FFFFFF}
\definecolor{CoreLightBlue}{HTML}{8BBDEB}
\colorlet{LightCoreGray}{CoreGray!20}
\colorlet{LightCoreBlack}{CoreBlack!20}
\colorlet{LightCoreGreen}{CoreGreen!30}
\colorlet{LightCoreRed}{CoreRed!20}
\colorlet{LightCoreYellow}{CoreYellow!20}
\colorlet{LightCoreWhite}{CoreWhite!20}

\definecolor{MyRed}{HTML}{e41a1c}
\definecolor{MyBlue}{HTML}{377eb8}
\definecolor{MyGreen}{HTML}{4daf4a}
\definecolor{MyPurple}{HTML}{984ea3}
\definecolor{MyOrange}{HTML}{ff7f00}
\definecolor{MyYellow}{HTML}{ffff33}
\definecolor{MyBrown}{HTML}{a65628}

\usepackage[absolute,showboxes]{textpos}
\usepgfplotslibrary{external}


\usepackage{pifont}
\newcommand{\cmark}{\ding{51}}%
\newcommand{\xmark}{\ding{56}}%

\usepackage{xspace}
\newcommand{\etal}{\textit{et~al.}~}
\newcommand{\eg}{\textit{e.g.,}~}

\newcommand{\cf}{\textit{cf.,}~}

\newcommand{\one}{({\em i})\xspace}
\newcommand{\two}{({\em ii})\xspace}
\newcommand{\three}{({\em iii})\xspace}
\newcommand{\four}{({\em iv})\xspace}

\newcommand{\nan}{\textit{NaN}\xspace}

\begin{document}

\bstctlcite{my:BSTcontrol}
\title{\LARGE \bf
    Coordinating Cooperative Perception in Urban Air Mobility for Enhanced Environmental Awareness$^{*}$
}
\author{Timo H{\"a}ckel$^{1}$, Luca von Roenn$^{2}$, Nemo Juchmann$^{3}$, \\Alexander Fay$^{2}$, Rinie Akkermans$^{3}$, Tim Tiedemann$^{1}$, and Thomas C. Schmidt$^{1}$%
\thanks{$^{*}$This work was funded by the German Hamburg X initiative within the i-LUM project [LFF-HHX-04.2] for innovative airborne urban mobility.}%
\thanks{$^{1}$Timo H{\"a}ckel (corresponding author \{\href{mailto:timo.haeckel@haw-hamburg.de}{timo.haeckel@haw-hamburg.de}\}), Tim Tiedemann, and Thomas C. Schmidt are with the Dept. of Computer Science, Hamburg University of Applied Sciences, Hamburg, Germany}%
\thanks{$^{2}$Luca von Roenn, and Alexander Fay are with the Institute of Automation Technology, Helmut Schmidt University, Hamburg, Germany}%
\thanks{$^{3}$Nemo Juchmann, and Rinie Akkermans are with the Dept. of Automotive and Aeronautical Engineering, Hamburg University of Applied Sciences, Hamburg, Germany}%
}

\maketitle
\thispagestyle{empty}
\pagestyle{empty}

\setlength{\TPHorizModule}{\paperwidth}
\setlength{\TPVertModule}{\paperheight}
\TPMargin{5pt}
\begin{textblock}{0.8}(0.1,0.02)
     \noindent
     \footnotesize
     If you cite this paper, please use the original reference:
     Timo H{\"a}ckel, Luca von Roenn, Nemo Juchmann, Alexander Fay, Rinie Akkermans, Tim Tiedemann, and Thomas C. Schmidt. 
     ``Coordinating Cooperative Perception in Urban Air Mobility for Enhanced Environmental Awareness,'' In: \emph{2024 International Conference on Unmanned Aircraft Systems (ICUAS)}. IEEE, June 2024.
\end{textblock}

\begin{abstract}
    The trend for Urban Air Mobility (UAM) is growing with prospective air taxis, parcel deliverers, and medical and industrial services. 
    Safe and efficient UAM operation relies on timely communication and reliable data exchange. 
    In this paper, we explore Cooperative Perception (CP) for Unmanned Aircraft Systems (UAS), considering the unique communication needs involving high dynamics and a large number of UAS. 
    We propose a hybrid approach combining local broadcast with a central CP service, inspired by centrally managed U-space and broadcast mechanisms from automotive and aviation domains.
    In a simulation study, we show that our approach significantly enhances the environmental awareness for UAS compared to fully distributed approaches, with an increased communication channel load, which we also evaluate. 
    These findings prompt a discussion on communication strategies for CP in UAM and the potential of a centralized CP service in future research. 
\end{abstract}

\begin{acronym}
	\acro{A2A}[A2A]{Air-to-Air}
	\acro{A2G}[A2G]{Air-to-Ground}
	\acro{ACC}[ACC]{Adaptive Cruise Control}
	\acro{ACDC}[ACDC]{Automotive Cyber Defense Center}
	\acro{ACL}[ACL]{Access Control List}
	\acro{AD}[AD]{Anomaly Detection}
	\acro{ADAS}[ADAS]{Advanced Driver Assistance System}
	\acro{ADS}[ADS]{Anomaly Detection System}
	\acroplural{ADS}[ADSs]{Anomaly Detection Systems}
	\acro{ADS-B}[ADS-B]{Automatic Dependent Surveillance-Broadcast}
	\acro{API}[API]{Application Programming Interface}
	\acro{AUTOSAR}[AUTOSAR]{AUTomotive Open System ARchitecture}
	\acro{AVB}[AVB]{Audio Video Bridging}
	\acro{ARP}[ARP]{Address Resolution Protocol}
	\acro{ATS}[ATS]{Asynchronous Traffic Shaping}
	\acro{BE}[BE]{Best-Effort}
	\acro{C2X}[Car2X]{Car-to-Everything}
	\acroplural{CA}[CAs]{Certification Authorities}
	\acro{CAN}[CAN]{Controller Area Network}
	\acro{CA}[CA]{Cooperative Awareness}
	\acro{CAM}[CAM]{Cooperative Awareness Message}
	\acro{CAS}[CAS]{Cooperative Awareness Service}
	\acro{CBM}[CBM]{Credit Based Metering}
	\acro{CBS}[CBS]{Credit Based Shaping}
	\acro{CNC}[CNC]{Central Network Controller}
	\acro{CMI}[CMI]{Class Measurement Interval}
	\acro{CP}[CP]{Cooperative Perception}
	\acro{CPM}[CPM]{Collective Perception Message}
	\acro{CPS}[CPS]{Collective Perception Service}
	\acro{CoRE}[CoRE]{Communication over Realtime Ethernet}
	\acro{CT}[CT]{Cross Traffic}
	\acro{CM}[CM]{Communication Matrix}
	\acro{DANE}[DANE]{DNS-Based Authentication of Named Entities}
 	\acro{DENM}[DENM]{Decentralized Environmental Notification Message}
	\acro{DENS}[DENS]{Decentralized Environmental Notification Service}
	\acro{DoS}[DoS]{Denial of Service}
	\acro{DDoS}[DDoS]{Distributed \acl{DoS}}
	\acro{DDS}[DDS]{Data Distribution Service}
	\acro{DNS}[DNS]{Domain Name System}
	\acro{DNSSEC}[DNSSEC]{Domain Name System Security Extensions}
	\acro{DPI}[DPI]{Deep Packet Inspection}
	\acro{E/E}[E/E]{Electrical/Electronic}
	\acro{EAR}[EAR]{Environment Awareness Ratio}
	\acro{ECU}[ECU]{Electronic Control Unit}
	\acroplural{ECU}[ECUs]{Electronic Control Units}
	\acro{ETSI}[ETSI]{European Telecommunications Standards Institute}
	\acro{FN}[FN]{False Negative}
	\acro{FP}[FP]{False Positive}
	\acro{FDTI}[FDTI]{Fault Detection Time Interval}
	\acro{FHTI}[FHTI]{Fault Handling Time Interval}
	\acro{FRTI}[FRTI]{Fault Reaction Time Interval}
	\acro{FTTI}[FTTI]{Fault Tolerant Time Interval}
	\acro{GCL}[GCL]{Gate Control List}
	\acro{GS}[GS]{Ground Station}
	\acroplural{GSs}[GS]{Ground Stations}
	\acro{HTTP}[HTTP]{Hypertext Transfer Protocol}
	\acro{HMI}[HMI]{Human-Machine Interface}
	\acro{HPC}[HPC]{High-Performance Controller}
	\acro{IA}[IA]{Industrial Automation}
	\acro{IAM}[IAM]{Identity- and Access Management}
	\acro{ICT}[ICT]{Information and Communication Technology}
	\acro{IDS}[IDS]{Intrusion Detection System}
	\acroplural{IDS}[IDSs]{Intrusion Detection Systems}
	\acro{IEEE}[IEEE]{Institute of Electrical and Electronics Engineers}
	\acro{IGMP}[IGMP]{Internet Group Management Protocol}
	\acro{IoT}[IoT]{Internet of Things}
	\acro{IP}[IP]{Internet Protocol}
	\acro{ITS}[ITS]{Intelligent Transport System}
	\acro{IVN}[IVN]{In-Vehicle Network}
	\acroplural{IVN}[IVNs]{In-Vehicle Networks}
	\acro{LIN}[LIN]{Local Interconnect Network}
	\acro{MAC}[MAC]{Message Authentication Code}
	\acro{MAVLink}[MAVLink]{Micro Air Vehicle Link}
	\acro{MTU}[MTU]{Maximum Transmission Unit}
	\acro{MOST}[MOST]{Media Oriented System Transport}
	\acro{NAD}[NAD]{Network Anomaly Detection}
	\acro{NADS}[NADS]{Network Anomaly Detection System}
	\acroplural{NADS}[NADSs]{Network Anomaly Detection Systems}
	\acro{OEM}[OEM]{Original Equipment Manufacturer}
	\acro{OMG}[OMG]{Object Management Group}
	\acro{OTA}[OTA]{Over-the-Air}
	\acro{P4}[P4]{Programming Protocol-independent Packet Processors}
	\acro{PCAP}[PCAP]{Packet Capture}
	\acro{PCAPNG}[PCAPNG]{PCAP Next Generation Dump File Format}
	\acro{PCP}[PCP]{Priority Code Point}
	\acro{PKI}[PKI]{Public Key Infrastructure}
	\acro{PSFP}[PSFP]{per-stream filtering and policing}
	\acro{RC}[RC]{Rate-Constrained}
	\acro{REST}[ReST]{Representational State Transfer}
	\acro{RPC}[RPC]{Remote Procedure Call}
	\acro{RSU}[RSU]{Road Side Unit}
	\acro{SD}[SD]{Service Discovery}
	\acro{SDN}[SDN]{Software-Defined Networking}
	\acro{SDN4CoRE}[SDN4CoRE]{Software-Defined Networking for Communication over Real-Time Ethernet}
	\acro{SOA}[SOA]{Service-Oriented Architecture}
	\acroplural{SOA}[SOAs]{Service-Oriented Architectures}
	\acro{SOME/IP}[SOME/IP]{\textit{Scalable service-Oriented MiddlewarE over IP}}
	\acro{SPOF}[SPOF]{Single Point of Failure}
	\acro{SR}[SR]{Stream Reservation}
	\acro{SRP}[SRP]{Stream Reservation Protocol}
	\acro{SUMO}[SUMO]{\textit{Simulation of Urban MObility}}
	\acro{SVM}[SVM]{Support Vector Machine}
	\acro{SW}[SW]{Switch}
	\acroplural{SW}[SWs]{Switches}
	\acro{TAS}[TAS]{Time-Aware Shaping}
	\acro{TCP}[TCP]{Transmission Control Protocol}
	\acro{TDMA}[TDMA]{Time Division Multiple Access}
	\acro{TN}[TN]{True Negative}
	\acro{TP}[TP]{True Positive}
	\acro{TLS}[TLS]{Transport Layer Security}
	\acro{TSN}[TSN]{Time-Sensitive Networking}
	\acroplural{TSN}[TSN]{Time-Sensitive Networks}
	\acro{TSSDN}[TSSDN]{Time-Sensitive Software-Defined Networking}
	\acro{TT}[TT]{Time-Triggered}
	\acro{TTE}[TTE]{Time-Triggered Ethernet}
	\acro{TTL}[TTL]{Time to Live}
	\acro{UAV}[UAV]{Unmanned Aerial Vehicle}
	\acro{UAS}[UAS]{Unmanned Aircraft System}
	\acroplural{UAS}[UAS]{Unmanned Aircraft Systems}
	\acro{UAM}[UAM]{Urban Air Mobility}
	\acro{UDP}[UDP]{User Datagram Protocol}
	\acro{UN}[UN]{United Nations}
	\acro{QoS}[QoS]{Quality-of-Service}
	\acro{V2X}[V2X]{Vehicle-to-X}
	\acro{WS}[WS]{Web Services}
	\acro{ZC}[ZC]{Zone-Controller}
	\acroplural{ZC}[ZCs]{Zone-Controllers}
\end{acronym}

\section{Introduction}%
\label{sec:introduction}

\ac{UAM} represents a prominent trend in aviation, aiming for efficient and safe aerial transportation within urban environments. 
It encompasses various applications including air taxis, package delivery services, medical transports, and industrial applications. 
\acp{UAS} play a crucial role in shaping the future of aerial transportation~\cite{gasy-aufjr-21}.

For (partially) autonomous \acp{UAS}, it is expected that ensuring environmental perception is fundamental for the safe and reliable operation of \ac{UAM}~\cite{yz-satjr-15, rhjgk-cdtjr-24}. 
The surveillance of static objects and potential conflicts within a large area becomes crucial for trajectory control, particularly at high speeds. 
A conflict occurs when two or more conflict elements overlap in space and time. 
Conflict elements may include objects like other airspace users, geofences, harsh weather conditions, or structures such as buildings~\cite{rgf-cadjr-23}. 
Local sensors alone remain insufficient for reliable conflict detection due to possible interruptions of the line-of-sight, limited sensing range, and disruptive effects introduced by the urban environment~\cite{slbv-cpsjr-20}. 
Therefore, effective communication is required to enable robust environmental awareness tailored to \acp{UAS}. 

Connectivity is already vital for various aspects within the U-space airspace -- a centrally managed airspace for \ac{UAS} operations~\cite{e-ussjr-22} -- such as mission planning and identity management. 
The U-space comprises diverse services including registration, network identification, and flight authorization. 
Cooperative obstacle detection is also envisioned but a unified data space structuring the available information to be shared and a suitable communication model are missing.

\ac{UAM} communication differs significantly from conventional aircraft, consumer drones, and ground vehicles. 
The unique characteristics of \ac{UAM}, including smaller separation, a denser environment, and proximity to urban factors including interfering with humans, and animals and quickly changing constraints, such as geofences, necessitate a specialized communication framework. 
The \ac{ADS-B}~\cite{r-adsb-06} enables receiving systems to locate the sending aircraft. 
Currently, information about the \ac{UAS} surroundings is not shared among aircraft. 
This contrasts the automotive sector, where \ac{V2X} communication has seen significant advancements~\cite{e-vccam-19,e-vccps-19,sbkl-apejr-20,sggmc-ampjr-22}. 
\acp{UAS} are usually operated by one person and only communicate with their operator. 
The centrally managed U-space further underscores the need for an efficient and standardized communication framework~\cite{e-ussjr-22}.

\ac{UAM} communication is still at an early stage, requiring exploration of use cases and information exchange~\cite{e-ussjr-22}. 
Investigation of combined sensing, \ac{CP}, and central U-space services are needed, and simulations can help with gaining insight into benefits and limitations. 
The lack of a unified and structured data space, a communication model, and standardized protocols motivate an investigation to unlock the full potential of \ac{UAS} communication. 

In this paper, we propose a hybrid approach to \ac{CP} combining broadcast with a central service for coordination of \ac{CP} in \ac{UAM}.
Therefore, we explore the \ac{CP} data space for UAS by comparing existing solutions across the automotive, drone, and aircraft domains, with a specific focus on the environmental perception of autonomous \acp{UAS}. 
We compare the performance of our hybrid approach to fully distributed mechanisms through simulations, focusing on the environmental awareness and the impact on the communication channel.
With this, we aim to initiate a discussion on communication requirements for \ac{CP} in \ac{UAM} and the advantages of adopting a centralized \ac{CP} service for future research. 

The remainder of this paper is structured as follows. 
Section~\ref{sec:background_and_related_work} provides an overview of related work. 
Section~\ref{sec:concept} presents our coordinated \ac{CP} model and identifies a \ac{CP} data space tailored for \ac{UAM}. 
In Section~\ref{sec:eval}, we investigate the performance of the proposed \ac{CP} service in a simulation scenario and compare to local perception and distributed approaches. 
Finally, Section~\ref{sec:conclusion_and_outlook} concludes the paper with an outlook on future work.

\section{Background and Related Work}
\label{sec:background_and_related_work}
\ac{UAS} in urban environments face an increasing risk of collisions due to reduced separation between \ac{UAS}, low altitudes, and the vicinity of urban influences. 
To efficiently detect and avoid conflicts, \acp{UAS} and the traffic management system require reliable environmental information, including traffic, geo data, weather, and geofences~\cite{rgf-cadjr-23}. 
Sensors alone cannot ensure safety in obstructed line-of-sight scenarios~\cite{slbv-cpsjr-20} and a limited detection range decreases the chance for early counteractions to possible collisions.

\ac{CA} and \ac{CP} facilitate real-time information sharing between \ac{UAS} and infrastructure, enhancing situational awareness, reducing the risk of accidents, and improving traffic flow. 
\ac{CA} focuses on exchanging information about the \ac{UAS} itself, such as position, speed, and direction, while \ac{CP} enables sharing environmental information, such as non-cooperative objects, extreme weather conditions, and animals (\eg flocks of birds). 
The following provides background and related work on the \mbox{U-space} and \ac{CP} in the aviation and automotive domains.

\subsection{Communication in \acl{UAM}}
Communication technology is essential for cooperative \acp{UAS}, enabling, for example, collaborative trajectory planning and collision avoidance~\cite{jsqfh-cccjr-23}. 
The \mbox{U-space} is a centrally managed airspace aimed at integrating and managing \acp{UAS}~\cite{gf-cusjr-22}, consisting of four development phases~\cite{e-ussjr-22}: 
Phase \one involves basic services, including registration, identification, and geofencing. 
Phase \two focuses on flight management services, such as trajectory planning, approval, and monitoring. 
Phase \three adds support for more complex scenarios, including automatic conflict resolution between interfering aircraft, obstacle detection, and avoidance. 
Lastly, phase \four represents the complete implementation of all \mbox{U-space} services, incorporating a high level of automation and interconnection between aircraft, pilots, authorities, and other stakeholders. 
Currently, \mbox{U-space} concepts and ideas for phases \one and \two are being implemented~\cite{gf-cusjr-22,e-ussjr-22}, and phases \three and \four are expected in 2027 and 2035~\cite{e-ussjr-22}.

\ac{UAM}-related communication protocols are still in early development. 
Multi-access Edge Computing (MEC) minimizes communication latencies for autonomous flight control~\cite{smsgq-maljr-23} and allows to offload detection algorithms to the mobile edge. 
Berling \etal ~\cite{bhwvg-muajr-21} investigate the number of messages exchanged between \acp{UAS} and \acp{GS} in an interdisciplinary toolchain that considers scenario generation, flight planning, and flight execution. 
Currently, there is a gap in analyzing environmental awareness with the introduction of a structured \ac{CP} data space for \ac{UAM}.

For efficient conflict detection and avoidance~\cite{rgf-cadjr-23, rhjgk-cdtjr-24} in \mbox{U-space} phase \three, communication-based \ac{CP} is necessary but not yet sufficiently evaluated. 
This paper collects a structured data space for distributed conflict detection and proposes a centralized \ac{CP} service for \ac{UAM}.

\subsection{Cooperative Awareness in the Aviation Domain}

Today, Air Traffic Control (ATC) relies on voice communication protocols between pilots and air traffic controllers, covering various topics such as flight planning, clearance delivery, departure, en route, and arrival procedures. 
However, voice communication does not apply to \acp{UAS}.

\ac{ADS-B}~\cite{r-adsb-06} is a surveillance technology broadcasting aircraft information to other aircraft and ground-based systems for traffic management and collision avoidance.
This information includes the position, heading, registration, mission, and type of the aircraft. 
Environmental information, such as weather conditions, obstacles, and non-cooperative objects, is not shared, as it is often unnecessary due to the large separation between vehicles. 
To the best of our knowledge, there are no specific protocols and standards for \ac{CP} in the aviation domain.

\subsection{Cooperative Perception in the Automotive Domain}

In contrast to the aviation domain, the automotive community has established standards for information exchange between vehicles and infrastructure, such as the ETSI \ac{CAS}~\cite{e-vccam-19}, and the evolving \ac{CPS}~\cite{e-vccps-19}. 
In particular, the \ac{CPS} defines messages for sharing data on detected objects, such as vehicles, pedestrians, or obstacles, including their position, velocity, heading, and other attributes. 
The term \textit{\acf{CP}} used in this paper further emphasizes the required collaboration between vehicles and infrastructure to obtain a comprehensive environmental picture.

The \ac{CPS} can improve environmental awareness for road vehicles~\cite{sbkl-apejr-20}. 
Still, current message generation rules may lead to detection latencies too high for safety-critical applications~\cite{sbkl-apejr-20}. 
Changing message generation can improve detection latency, but at the same time increases channel load and may cause channel instability~\cite{tsg-amgjr-19}. 
There is open potential for which objects are shared by which vehicles~\cite{sbkl-apejr-20} that help to reduce data redundancy and thus optimize channel utilization~\cite{hlssf-drmjr-20}.

In the automotive domain, common issues with \ac{CP} include the market penetration rate~\cite{slbv-cpsjr-20}, privacy concerns, and difficulties in defining a mission for private vehicles. 
These challenges do not translate to a centrally managed \mbox{U-space} airspace, where \ac{CP} can be enforced, and private vehicles are not expected to be part of \ac{UAM}. 
Nonetheless, concepts such as intent sharing and collaboration for trajectory optimization~\cite{asha-cpbjr-23} are also central features of \ac{UAM}. 

Drawing from the automotive experience and existing standards, the centralized \mbox{U-space} provides a unique opportunity to implement \ac{CP} in a scalable and efficient manner. 
Our analysis includes automotive protocols in the \ac{CP} data space, discussing their applicability to \ac{UAM} and highlighting domain differences and the potential for improved \ac{CP} efficiency with central management.

\section{Coordinating Cooperative Perception in Urban Air Mobility}%
\label{sec:concept}
During operation, \ac{UAS} need to be aware of their environment.
The combined local sensors provide environmental awareness for the direct vicinity of the \ac{UAS}. 
For safe reaction time at high speeds objects must be detected early, \eg object detection at \SI{500}{\meter} distance with a speed of \SI{50}{\meter\per\second} leaves only \SI{10}{\second} for evasive maneuvers.

\ac{UAS} can transmit various information to other \ac{UAS} or the infrastructure for sharing their current state, mission, or perception of the environment. 
\ac{CA} ensures that \ac{UAS} are aware of each other by sharing metadata, kinematics, and mission information about themselves that allow to identify the \ac{UAS} and predict future behavior. 
Beyond that, \ac{CP} enables \ac{UAS} to obtain a comprehensive environmental perception by sharing information about the environment, conflicts, and hazards.
Such data is intended to improve situational awareness --- particularly for \acp{UAS} --- and enable safer and more efficient operations. 

\ac{CP} in \ac{UAM} should allow interoperability between different parties, avoiding vendor lock-in using open standards. 
Crowded areas should be handled without overloading the channel, while at the same time, larger distances between \acp{UAS} should not lead to a lack of information. 
Additionally, the latency from object detection to sharing perception and receiving perception updates should be minimized to ensure data freshness.
In the following, we identify relevant information for \ac{UAM}, its freshness requirements, and compare appropriate distribution strategies.

\subsection{Cooperative Perception Model for Urban Air Mobility}
\label{sec:concept_cp_models}

\ac{UAS} will use different communication models to interact with their surroundings.
In a U-space, \ac{UAS} utilize various centralized services to register and identify themselves, plan their missions, and receive flight authorizations.
Conventional \ac{CA} and \ac{CP} systems, such as \ac{ADS-B}~\cite{r-adsb-06} are distributed, broadcasting information to all vehicles in communication range.
Drones utilize protocols such as \ac{MAVLink}~\cite{mavlink} for remote control and mission planning based on a closed peer-to-peer network of known parties limited in size (up to 255 concurrent systems). 

Combining the approaches is promising for \ac{UAM} to achieve central airspace control, distributed surveillance, and dedicated remote control for \ac{UAS}.
Central \mbox{U-space} services, however, commonly use peer-to-peer connections and do not benefit from \ac{CA} and \ac{CP} announcements. 
We propose a hybrid approach combining distributed and centralized elements using a limited broadcast range and long-range centralized management, thereby enhancing the environmental awareness for both \ac{UAS} and central \mbox{U-space} providers.

\begin{figure}
    \centering
    \includegraphics[width=1.0\linewidth]{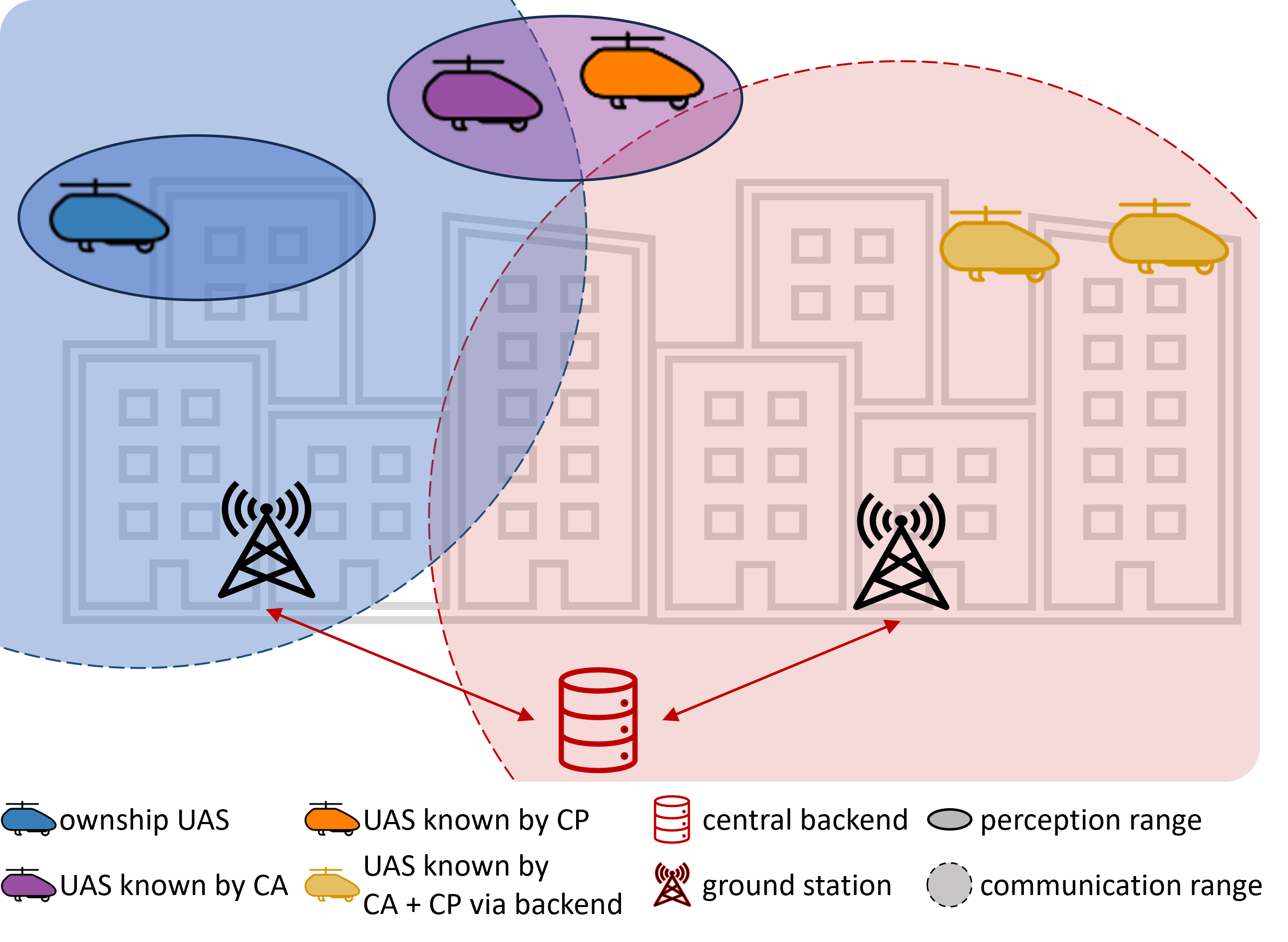}
    \caption{Hybrid CP model integrating local perception, CP broadcast, and a central backend service. Ground stations connect UAS in range to the backend.}
    \label{fig:communication}
\end{figure}

Fig.~\ref{fig:communication} shows an example of the hybrid \ac{CP} model for \ac{UAM}.
The ownship \ac{UAS} (blue) perceives its immediate surroundings through a combination of local sensors.
All \ac{UAS} broadcast their position and status (\ac{CA}) making the ownship \ac{UAS} aware of the position of the purple \ac{UAS}. 
In addition, \ac{UAS} broadcast objects detected with their sensors (\ac{CP}). 
The ownship \ac{UAS} detects the orange \ac{UAS} through a message from the purple \ac{UAS}.

Beyond the local broadcast (\ac{CA} and \ac{CP}), \acp{GS} function as gateways forwarding received messages broadcasts from the \ac{UAS} to a central \ac{CP} service in the backend. 
The backend caches and aggregates this information to track \ac{UAS} and other detected objects across the urban area.
The backend redistributes this aggregated information to all \ac{GS} via unicast. 
Reducing the redistributed information to a region of interest around the \ac{GS} can reduce the channel load.
The \acp{GS} provide the aggregated information to the \ac{UAS} again using \ac{CP} broadcast, which extends the detection range for all \ac{UAS} in the communication range of \acp{GS}.
In the example, the backend tells the ownship \ac{UAS} about the two yellow \ac{UAS}, providing complete environmental awareness. 
This supports functions such as automated trajectory control, conflict detection, and resolution mechanisms.

\subsection{Cooperative Perception Data Space}%

In \ac{UAM}, existing aviation standards, such as \ac{ADS-B}~\cite{r-adsb-06}, lack comprehensive information and cooperative maneuvers for \acp{UAS}. 
Automotive standards such as the ETSI \ac{CAS}~\cite{e-vccam-19} and the evolving \ac{CPS}~\cite{e-vccps-19} offer information exchange adaptable to \acp{UAS}. 
Innovations such as \ac{MAVLink} further enhance drone technology.

Table~\ref{tab:vehicle_information} compiles a data space from standards, protocols, and the state-of-the-art in the automotive, aviation, and drone domains. 
In our analysis, we focus on the \ac{CP} use case for \ac{UAM}. 
Additional services including weather data, mission planning, remote control, and direct \acp{UAS} communication, \eg for conflict resolution, are out of the scope of this work.

Metadata encompasses the \ac{UAS} type, dimensions, and operational status (\eg on a mission, take-off, or landing). 
Special vehicles, such as rescue helicopters, may inform their environment about an ongoing operation. 
A public ID is used for registration and network identification in a \mbox{U-space}. 
The current time and a time delta are shared to determine the age of information and synchronize the local clocks.

Kinematic information includes the position, velocity, acceleration, and speed of the \ac{UAS}. 
Sharing heading or orientation helps to determine the intended direction of travel. 
In addition, \ac{UAS} may communicate their intent~\cite{asha-cpbjr-23} with mission information, such as the current plan, status, and changes from the planned route, which is essential for \mbox{U-space} operations. 
This includes transmitting information related to the mission or destination, such as delivery schedules or passenger pickup/drop-off locations. 

\acp{UAS} employ sensors (cameras, LIDARs, RADARs) and algorithms (\eg based on machine learning) for object detection and classification. 
By sharing detected objects, sensing capabilities (\eg field of view, range, and resolution), and confidence in classification, \ac{UAS} can extend the perceived environment to detect and resolve conflict elements along the planned trajectory early on.

For road vehicles, information about traffic conditions, such as congestion, accidents, road closures, and empty road space helps vehicles plan their routes. 
This can also be applied to \ac{UAS} operations, communicating obstacles on flight routes. 
Traffic signals and guidance convey \ac{UAM} traffic, sector capacity, and geofences. 

In the urban environment, not only \acp{UAS} can cause conflicts but also other hazards, such as unknown or non-cooperative \ac{UAS}, construction sites, animals (\eg flocks of birds), or extreme weather conditions (\eg wind, rain, fog). 
A central database cashes and tracks these conflict elements across the urban area and helps to identify conflicts along the trajectory of the \ac{UAS}. 

Besides safety, the sound emission caused by \acp{UAS} is one of the main concerns of the public regarding \ac{UAM}~\cite{eu-dsuas-22}. 
For noise reduction purposes, it is advisable that \ac{UAS} record and process acoustic signals during operation. 
A wireless acoustic sensor network is suitable for this context~\cite{ahaa-msnnm-17, ar-enmsc-19}. 
By communicating the emitted sound based on practical metrics, \eg the Sound Pressure Level (SPL) or Effective Perceived Noise Levels (EPNL)\cite{Smith_1989}, the central backend can create a noise map for the urban area. 
Areas with high noise can then be treated as a conflict. 

Conflicts can affect \ac{UAS} operation and require increased separation, for example. 
Detected conflicts are shared to avoid collisions, including \ac{UAS} breakdowns, traffic rule violations, and hazards.

\begin{table}
    \caption{
        \ac{UAM} data space sourced from various automotive, aircraft, and drone protocols.
        Shows information that is shared [\cmark], not shared [\xmark] in the respective domain, or shared but not sufficient [(\cmark)] for UAS.
        The frequency describes how often the data is included in generated messages and is based on the highest in any industry. 
        The frequency of NaN is not specified in a standard or not considered for this work.
    }
    \label{tab:vehicle_information}
    \setlength{\tabcolsep}{2pt}  
    \begin{tabularx}{\linewidth}{p{0.35cm} l c c c c} 
        \toprule
        \multicolumn{2}{l}{\textbf{Vehicle Information}} 
        & \textbf{Automotive}
        & \textbf{Aviation}
        & \textbf{Drone}
        & \textbf{Frequency} $[Hz]$ \\
        
        \midrule
        \multirow{6}{*}{\rotatebox{90}{Metadata}}
        & Current time & \cmark & \cmark & \cmark & every Msg\\
        & Time delta & \cmark & \xmark & \cmark & every Msg\\
        & Type, dimensions & \cmark & \cmark & \cmark & 2 \\
        & Special vehicle & \cmark & \xmark & \xmark & 2 \\
        & Public ID & \cmark & \cmark & \cmark & 2 \\
        & Vehicle status & (\cmark) & \cmark & \cmark & 2 \\
        
        \midrule
        \multirow{5}{*}{\rotatebox{90}{Kinematic}}
        & Position & \cmark & \cmark & \cmark & 1 to 10 \\
        & Altitude & \xmark & \cmark & \cmark & 1 to 10 \\
        & Heading, velocity & \cmark & \cmark & \cmark & 1 to 10 \\
        & Acceleration & \cmark & \cmark & \cmark & 1 to 10 \\
        & Remote control & \xmark & \xmark & \cmark & \nan \\

        \midrule
        \multirow{5}{*}{\rotatebox{90}{Mission}}
        & Service provider & \xmark & (\cmark) & \cmark & 2 \\
        & Public mission ID & \xmark & \cmark & \xmark & 2 \\
        & Plan & \xmark & \xmark & \cmark & on request / 2 \\
        & Mission status & \xmark & \xmark & \cmark & on request / 2 \\
        & Changes & \xmark & \xmark & \cmark & \nan \\

        \midrule
        \multirow{5}{*}{\rotatebox{90}{Environment}}
        & Objects, vehicles & \cmark & \xmark & \cmark & 1 to 10 \\
        & Traffic guidance & \cmark & \xmark & \xmark & 1 to 10 \\
        & Weather, wind & \xmark & (\cmark) & (\cmark) & \nan \\
        & Noise & \xmark & \xmark & \xmark & 1 to 10 \\
        & Stream sensor data & \xmark & \xmark & \cmark & \nan \\
        
        \midrule
        \multirow{4}{*}{\rotatebox{90}{Conflict}}
        & Hazards & \cmark & \xmark & (\cmark) & 1 to 10 \\
        & Collision avoidance & (\cmark) & (\cmark) & \cmark & 1 to 10 \\
        & Vehicle break down & \cmark & \xmark & \cmark & 1 to 10 \\
        & Traffic rule violation & \cmark & \xmark & \xmark & 1 to 10 \\
        \bottomrule
    \end{tabularx}
\end{table}

\subsection{Information Lifetime and Frequency}
The frequency of information exchange varies depending on the specific use case and application. 
It is determined by factors, including the required level of situational awareness, the lifetime of the data object, the communication range and bandwidth of the wireless network, and the processing capabilities of the \ac{UAS} and infrastructure involved. 
In general, \ac{UAS} should reduce the number of transmissions to save resources, but the data must be fresh enough to be useful, especially for fast-moving objects.

Conventional aircraft use lower frequencies due to larger inter-vehicle distances and stable environments. 
In contrast, \ac{UAM} demands higher frequencies for closer proximity and dynamic conditions. 
Ground vehicles typically use higher frequencies, reflecting their closer spacing and dynamic environment. 
Some standards adjust transmission frequency based on vehicle speed and channel congestion~\cite{e-vccam-19}.

Table~\ref{tab:vehicle_information} presents the highest sharing frequency among industries considered, representing the most demanding use case. 
Metadata and mission information rarely change and are shared twice per second to allow identification of \ac{UAS}.

\ac{UAS} share their kinematics and speed at least once and up to ten times per second depending on the \ac{UAS} dynamics and the channel congestion status. 
\ac{ADS-B} requires a message every \SI{5}{\second} on the ground and twice per second when in the air. 
ETSI \ac{CA} messages are triggered when the last message was sent more than \SI{1}{\second} ago, or when the current vehicle heading differs from the previously announced by more than \SI{4}{\degree}, the position differs by more than \SI{4}{\meter}, and the speed differs by more than \SI{0.5}{\meter\per\second}. 
This could be extended to include altitude changes.

\ac{CP} is not established in the aircraft domain and detected objects are not communicated as it is often unnecessary due to the large separation between vehicles. 
Instead established voice communication protocols are used. 
In the automotive sector, the \ac{CPS}~\cite{e-vccps-19} has attracted a lot of interest, so we base our assessment on the experience gained there.
Information on detected objects is shared with a frequency of \SIrange{1}{10}{\hertz} depending on changes in their kinematics following the same rule set as \ac{CA} messages. 

\acp{GS} also share an aggregated view of detected moving objects in an area of interest at least once per second and up to \SI{10}{\hertz}. 
Traffic signals and guidance will also follow such rules, for changing geofences, traffic conditions, noise pollution, and sector capacity, making sure at least one message is sent while a \ac{UAS} is in range. 
Static objects can be requested by the \ac{UAS} at once when entering a new area, with deviations reported promptly.

\subsection{Communication Model}
\label{sec:communication_model}
Various standards coexist for vehicular communication. 
For instance, \ac{ADS-B} broadcasts at a lower radio frequency (around \SI{1}{\giga\hertz}) that can be received at a long distance especially for high-altitude flights, sacrificing transfer speed and data throughput. 
For road vehicles, different standards have prevailed in different regions, for example, large-scale deployments of the \mbox{IEEE 802.11p} WLAN-based \ac{V2X} in the EU and Cellular-\ac{V2X} (based on LTE and 5G) in the US with still few deployments. 
Both of which utilize higher radio frequencies (\SI{2.4}{\giga\hertz} or \SI{5}{\giga\hertz}) allowing for larger data transfers at decreased range (a few \SI{100}{\meter}). 

Each of the different technologies has its advantages and disadvantages. 
Using long-range broadcast for the numerous \acp{UAS} in a \mbox{U-space} airspace causes a significant increase in message volume, leading to potential channel congestion and interference with conventional aircraft. 
Limiting the broadcast range can reduce the congestion but also limits the number of \ac{UAS} that can be reached directly.
On the other hand, higher frequencies can allow for a larger data throughput, which will be required for large \ac{CP} data. 

To enable both direct interaction between \ac{UAS} and communication with \mbox{U-space} services, the \ac{UAM} communication system should support both direct device-to-device communication and Internet connectivity. 
Therefore, Cellular-\ac{V2X} adds a direct device-to-device communication mode and WLAN-based \ac{V2X} can use infrastructure as a gateway to \mbox{U-space} services.
Regardless of the technology chosen, the cooperative perception model should be independent of the communication technology, as exemplified by ETSI \mbox{ITS-G5} standards for intelligent road vehicles. 

In our evaluation, we use \mbox{IEEE 802.11p} and the \mbox{ITS-G5} protocols to communicate \ac{CA} and \ac{CP} messages to other \ac{UAS} and via infrastructure to the backend. 
For a detailed comparison of IEEE \mbox{802.11p} and \mbox{LTE-V2X}, we refer to~\cite{mgs-ci8jr-20}.
We compare local perception without communication and the distributed \ac{CA} and \ac{CP} approaches for vehicles with our hybrid approach with centralized collection and distribution of \ac{CP} data using a central backend entity. 
We focus on environmental awareness and side effects introduced by the centralized \ac{CP} service, which should be unaffected by the chosen communication system.

\section{Case Study of Cooperative Perception in Urban Air Mobility}
\label{sec:eval}
We now evaluate the CP models (\cf Sec.~\ref{sec:concept_cp_models}) in a simulation case study comparing detection performance and communication metrics, along with assessing the impact of a central backend for \ac{CP}. 
We utilize the \ac{EAR} as a common metric of perception performance~\cite{slbv-cpsjr-20} that represents the ratio between detected and existing objects in the simulation. 
Additionally, we investigate features such as the detection delay and the impact of caching in a central backend.
For our communication analysis, we consider the number of messages, payload length, and receive (rx) channel load of the \ac{UAS}.

\subsection{Simulation Environment}
We use OMNeT++ with the INET framework~\cite{omnet-inet} and Artery~\cite{rog-alsjr-19} for \ac{V2X} communication simulations based on IEEE \mbox{802.11} and \mbox{ITS-G5} protocols. 
For a detailed evaluation of the \mbox{802.11p}-based communication between vehicles we refer to~\cite{sggmc-ampjr-22}.
Although we use WLAN-based communication (\SI{5.9}{\giga\hertz}), our results for environmental awareness directly apply to other communication technologies with similar range and bandwidth such as 5G or LTE \mbox{C-V2X}.

Unlike other \ac{V2X} simulators, Artery offers an environment model that allows simulating sensor perception of vehicles. 
Recent work improved sensor simulation with more realistic models~\cite{wygw-tirjr-23}, but we stick to the original sensor models as they are sufficient for the evaluation of the different perception models for \ac{UAS}. 
For our evaluations, we extend Artery with our centralized perception service in \acp{GS} and a backend node.

Artery uses the \ac{SUMO} traffic simulator~\cite{sumo} for vehicle simulation on a road network, but it cannot simulate \acp{UAS} yet. 
Efforts for a \ac{UAS} traffic simulation similar to \ac{SUMO}~\cite{hlps-tosjr-22} exist but are not integrated with Artery. 
A toolchain for \ac{UAM} simulations is shown by Berling \etal \cite{bhwvg-muajr-21} but is not publicly available.
Our goal for future work is to integrate the used simulation environment with other simulators, \eg for demand modeling and trajectory planning, in an RCE~\footnote{RCE integration environment: \url{https://rcenvironment.de/}} toolchain to enable a realistic evaluation of \ac{UAS} perception models.

\ac{UAM} aircraft, traffic guidelines, and regulations are still in active development and vary significantly between regions. 
Currently, the \mbox{U-space} and \ac{UAM} are not deployed at a large scale which limits the availability of realistic models that include demand, trajectories, \ac{UAS} behavior, vertiport locations, and geofences. 
The use of Artery and \ac{SUMO} confines \ac{UAS} and sensor simulation to two dimensions, projecting flight trajectories to ground level. 
This does not impact the communication behavior, which is modeled in all three dimensions but simplifies the sensor simulation. 
While sufficient for our comparison of the perception models for \ac{UAS}, future work should extend simulations to three dimensions, considering \ac{UAS} mobility modeled in all directions.

\subsection{Scenarios}
\begin{figure}[t]
	\centering
	\begin{minipage}{\linewidth}
		\centering
		\includegraphics[width=0.93\linewidth]{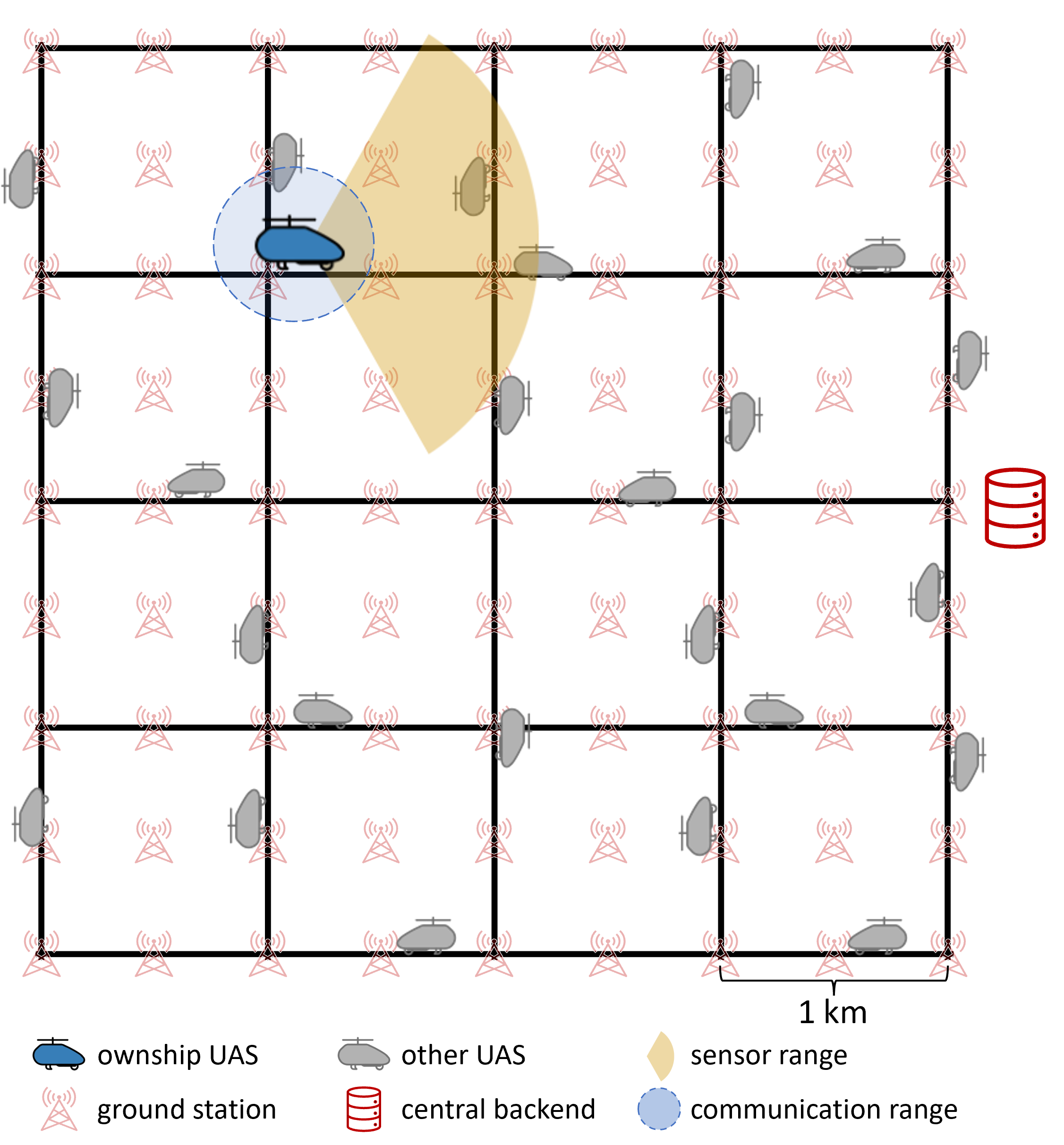}
		\caption{Evaluation setup consisting of 200 \ac{UAS} randomly distributed on a \SI{4}{\kilo\meter}-by-\SI{4}{\kilo\meter} grid network. Only the scenario w/ central backend adds \acp{GS} that cover the entire area.}
		\label{fig:scenario}		
	\end{minipage}
	\vspace{4pt}

	\begin{minipage}{\linewidth}
		    \begin{tikzpicture}
        \begin{groupplot}[
            width=0.7\linewidth,
            height=0.225\linewidth,
            group style={
                group size=1 by 3,
                horizontal sep = 0pt, 
                vertical sep = 3pt,
                xlabels at = edge bottom,
                xticklabels at = edge bottom, 
                ylabels at = edge left,
                yticklabels at = edge left, 
            },
            xlabel={Ground stations [\#]},
            xmajorgrids=true,
            scale only axis,
            every outer y axis line/.append style={black},
            ytick pos=left,
            xmin=0, xmax=225,
            xtick={0,25,36,49,64,81,100,121,144,169,196,225},
            xticklabels={0,25,36,49,64,81,100,121,144,169,196,225},
            x tick label style={rotate=90,anchor=east},
            xlabel shift=-2pt,
        ]
        \nextgroupplot[
            ylabel style = {align=center, text width=1cm},
            ylabel={Avg EAR},
            ymin=-0.05, ymax=1.05,
            ytick={0,0.5,1},
            yticklabel={\pgfmathparse{\tick*100}\pgfmathprintnumber{\pgfmathresult}\%},
            ylabel shift=-7pt,
            legend pos=south east,
        ]
        \addplot [thick, no markers, MyGreen] table [x=groundStations, y=earAvgNodes, col sep=comma] {rsu_study_ear.csv} ;\label{tikz:ear_nodes}
        \addplot [thick, no markers, MyRed] table [x=groundStations, y=earAvgBackend, col sep=comma] {rsu_study_ear.csv} ;\label{tikz:ear_backend}

        \nextgroupplot[
            ylabel style = {align=center, text width=2cm},
            ylabel={Avg rx channel load},
            ymin=-0.01, ymax=0.25,
            ytick={0,0.1,0.2},
            ylabel shift=-2pt,
            yticklabel={\pgfmathparse{\tick*100}\pgfmathprintnumber{\pgfmathresult}\%},
            legend columns=4,
            legend style = {
                rotate=90,
                rounded corners,
                font=\footnotesize,
                at={(1,2.33)},
                },
        ]
        \addplot [thick, no markers, MyGreen] table [x=groundStations, y=channelLoadAvg, col sep=comma] {rsu_study_channelload.csv};\label{tikz:channelload}
        
        \nextgroupplot[
            ylabel style = {align=center, text width=2cm},
            ylabel={Avg message count},
            ylabel shift=-7pt,
            legend columns=3,
            ymax=1100,
            legend pos=north east,
        ]
        \addplot [thick, no markers, MyOrange] table [x=groundStations, y=messagesRsuAvg, col sep=comma] {rsu_study_messageCount.csv}; \label{tikz:messagecount_rsu}
        \addplot [thick, no markers, MyBlue] table [x=groundStations, y=messagesTotalAvg, col sep=comma] {rsu_study_messageCount.csv}; \label{tikz:messagecount_all}
        \addplot [thick, no markers, MyGreen] table [x=groundStations, y=messagesNodesAvg, col sep=comma] {rsu_study_messageCount.csv}; \label{tikz:messagecount_nodes}
        \end{groupplot}
        
        \draw[draw=black, rounded corners] (6.2,-4.0) rectangle ++(0.4,5.9)node[midway,rotate=90, font=\footnotesize]{UAS \ref{tikz:ear_nodes} GS \ref{tikz:messagecount_rsu} Backend \ref{tikz:ear_backend} Total \ref{tikz:messagecount_all}};
    \end{tikzpicture}
    \vspace{-4pt}
    \caption{
        Impact of the number of ground stations on the average EAR, channel load, and message count. 
        The ground stations are equally spaced along the grid.
    }
    \label{fig:rsustudy}    
	\end{minipage} 
	\vspace{-15pt}
\end{figure}

Fig.~\ref{fig:scenario} shows the \SI{4}{\kilo\meter}-by-\SI{4}{\kilo\meter} simulation area in a grid network. 
\acp{UAS} move on predefined routes at a maximum speed of \SI{70}{\meter\per\second}. 
During the first \SI{20}{\second}, 200 \acp{UAS} spawn randomly distributed on the grid and despawn when they reach their destination. 
Each \ac{UAS} is equipped with one front RADAR sensor, with a range of \SI{1}{\kilo\meter} and a \SI{120}{\degree} field of view for precise object detection in front of the \ac{UAS}. 
With this, we simulate a large range with perfect detection capabilities. 
Specific sensor characteristics or sensor fusion techniques for near and far object detection are out of the scope of this work.
The communication range of the \acp{UAS} and \acp{GS} depends on the interference and path loss and is determined by the transmitter power. 
We use common settings for our parameters such as a transmitter power of \SI{200}{\milli\watt}, and a data rate of \SI{6}{\mega\bit\per\second}~\cite{sggmc-ampjr-22}. 
This results in a maximum range of several hundred meters.
The backend and \acp{GS} are connected using high-speed (\SI{100}{\giga\bit\per\second}) wired connections to eliminate a potential bottleneck.
Message send frequency between \SIrange{1}{10}{\hertz} is determined during runtime by the \mbox{ITS-G5} stack and depends on the channel load.
For this evaluation, we focus on the transmission of \ac{UAS} metadata and kinematics via \ac{CA} and \ac{CP} messages.

We simulate five coordination models for CP discussed in Sec.~\ref{sec:concept_cp_models} for \SI{100}{\second} of simulation time.
The scenarios follow the visualization of Fig.~\ref{fig:communication}:
\begin{enumerate}
	\item \textbf{Local perception}: \ac{UAS} use only local sensors without communication.
	\item \textbf{CA}: \ac{UAS} share their metadata and kinematics.
	\item \textbf{CP}: \ac{UAS} share information about detected objects.
	\item \textbf{CA \& CP}: \ac{UAS} share information about themselves and detected objects.
	\item \textbf{w/ central backend}: \ac{UAS} share information about themselves and detected objects. \acp{GS} covering the \SI{16}{\square\kilo\meter} simulation area forward incoming messages to a central backend service, which aggregates the perception of all \ac{UAS} and sends aggregated \ac{CP} messages to all \ac{UAS} via the \acp{GS}.
\end{enumerate}

\subsection{Results}

First, we determine the number of \acp{GS} used for the scenario w/ central backend. 
Fig.~\ref{fig:rsustudy} shows the average \ac{EAR} for \ac{UAS} and the backend, the average rx channel load, and the average number of messages per \ac{UAS}, \ac{GS}, and in total.
The number of \acp{GS} is increased from 0 to 225 and equally distributed on the grid.
For example, when five \acp{GS} are placed per row and column, the total number of \acp{GS} is 25.
The backend EAR and message count per \ac{GS} are only available if a \ac{GS} is placed.

The average channel load increases with the number of \acp{GS} from 17\% to 21\%.
The average \ac{EAR} for the backend is around 100\% already with 25 \acp{GS}.
The average \ac{EAR} for the \ac{UAS} improves drastically when adding \acp{GS}, reaching 66\% with 81 \acp{GS}.
After that, it does not increase significantly and even decreases by about 15\% when adding more \acp{GS} until it rises again.
This can be explained when looking at the number of messages sent.
\ac{UAS} and \acp{GS} reduce the number of messages sent when the channel load increases, which is the default behavior of the utilized \mbox{ITS-G5} stack to reduce channel congestion.
The optimal range and frequency of \ac{CP} messages and the ideal number and placement of \acp{GS} remain open research questions for future work.

In the following analyses, we use 81 \acp{GS} with a spacing of \SI{500}{\meter} for the scenario w/ central backend.
Table~\ref{tab:communication_results} shows the condensed results for the \ac{EAR}, payload length, and channel load for the five scenarios.
The channel load in the local perception scenario is caused by simulated noise that is also present in all other scenarios.
The minimum \ac{EAR} is zero in all cases because no objects are detected in the beginning. 

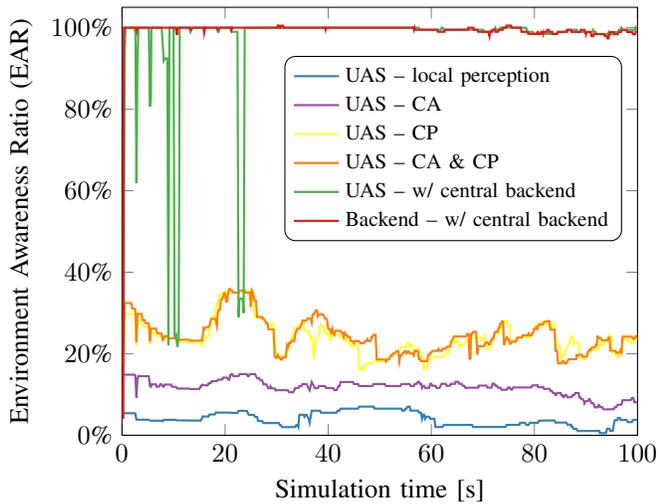
\begin{figure}
	    \centering
    \begin{tikzpicture}
        \begin{axis}[
            xlabel={Simulation time [s]},
            ylabel={Environment Awareness Ratio (EAR)},
            yticklabel={\pgfmathparse{\tick*100}\pgfmathprintnumber{\pgfmathresult}\%},
            xmin=0, xmax=100,
            ymin=0, ymax=1.05,
            legend style={
                rounded corners,
                font=\footnotesize,
                at={(0.97,0.67)}, 
                legend cell align={left},
                anchor=east, 
                legend columns=1, 
            },
        ]
        \addplot [thick, no markers, MyBlue] table [x=time, y=value, col sep=comma] {node51_sensing_ear.csv};
        \addplot [thick, no markers, MyPurple] table [x=time, y=value, col sep=comma] {node51_sensing+ca_ear.csv};
        \addplot [thick, no markers, MyYellow] table [x=time, y=value, col sep=comma] {node51_sensing+cp_ear.csv};
        \addplot [thick, no markers, MyOrange] table [x=time, y=value, col sep=comma] {node51_sensing+ca+cp_ear.csv};
        \addplot [thick, no markers, MyGreen] table [x=time, y=value, col sep=comma] {node51_sensing+ca+cp+backend_ear_81.csv};
        \addplot [thick, no markers, MyRed] table [x=time, y=value, col sep=comma] {backend_sensing+ca+cp+backend_ear_81.csv};
        \legend{
            UAS -- local perception,
            UAS -- CA,
            UAS -- CP,
            UAS -- CA \& CP,
            UAS -- w/ central backend,
            Backend -- w/ central backend
            }
        \end{axis}
    \end{tikzpicture}
    \caption{
        Environment Awareness Ratio (EAR) for one UAS and the backend with different communication enabled. In the scenario w/ central backend, 81 ground stations are used.
    }
    \label{fig:earplot}
\end{figure}

The \ac{EAR} over time is shown in Fig.~\ref{fig:earplot} for the perception models.
Local perception has the lowest \ac{EAR} as it depends on sensor range, which may vary for different sensor technologies.
Adding communication increases the \ac{EAR}.
\ac{CA} extends the range for detecting cooperative \acp{UAS}, however, with a small range, a lot of \acp{UAS} are already detected with local sensors. 
Noteworthily, \ac{UAS} sharing their detection (\ac{CP}) increases the \ac{EAR} significantly compared to \ac{CA}.
Furthermore, \ac{CP} adds the capability of sharing information about uncooperative objects.
Combining \ac{CA} and \ac{CP} does not improve \ac{EAR} compared to \ac{CP} alone; in some cases, it may even reduce it due to increased channel congestion from higher message volume.
Integrating \acp{GS} to relay \ac{CA} and \ac{CP} messages to a central perception service expands the perceivable area to 100\%, demonstrating the substantial benefit of central coordination, even in a small \SI{16}{\square\kilo\meter} area. 

The payload of \ac{CP} messages is significantly larger than \ac{CA} messages, which is reflected in the channel load (\cf Table~\ref{tab:communication_results}).
Furthermore, the payload length of \ac{CP} messages increases with the number of detected objects.
This is especially the case for the scenario w/ central backend, where the backend aggregates the perception of all \ac{UAS} and sends aggregated \ac{CP} messages to all \ac{UAS} via the \acp{GS}.
The amount of data included in the \ac{CP} messages should be carefully considered, as it directly impacts the channel load and transmission times, which in turn can lead to a reduced data freshness.
Scaling to larger urban areas increases data and communication volume, suggesting the need for optimization strategies of the central perception model. 
These could include private unicast or restricting \ac{GS} \ac{CP} content to conflict elements in a specific range.

Fig.~\ref{fig:earplot} also shows that although the average \ac{EAR} for the \ac{UAS} improves significantly with central coordination, it also decreases when no messages from the backend are received, \eg due to congestion, and cached objects timeout.
Still, the \ac{EAR} stays at the level of the best case for the other scenarios, because of the hybrid approach combining local perception, broadcast communication, and central coordination.
The \ac{EAR} rises again when the next message arrives.

\begin{figure}
	    \centering
    \begin{tikzpicture}[spy using overlays={circle, magnification=8, size=2cm, connect spies}]
        \begin{axis}[
            xlabel={Simulation time [s]},
            ylabel={Known UAS [\#]}, 
            xmin=0, xmax=100,
            ymin=0, ymax=210,
            legend style={
                rounded corners,
                font=\footnotesize,
                at={(0.97,0.16)}, 
                legend cell align={left},
                anchor=east, 
                legend columns=1, 
            },
        ]
        \addplot [no markers, MyRed, very thin,forget plot] table [x=time, y=value, col sep=comma] {backend_sensing+ca+cp+backend_81_vehiclecount.csv};
        \addplot [no markers, MyGreen, very thin,forget plot] table [x=time, y=value, col sep=comma] {node51_sensing+ca+cp+backend_81_vehiclecount.csv};
        \addplot [no markers, MyBlue, very thin, forget plot] table [x=time, y=value, col sep=comma] {world_sensing+ca+cp+backend_81_vehiclecount.csv};
        \spy[MyOrange] on (1.3,5.35) in node at (1.9,3.6); 
        \node[fill=none, anchor=north] at (27.8,95) {detection delay};
        \spy[MyPurple] on (2.1,5.4) in node at (5.6,3.6); 
        \node[fill=none, anchor=north] at (81.8,95) {cache eviction};
        \addlegendimage{MyRed, thick}
        \addlegendimage{MyGreen, thick}
        \addlegendimage{MyBlue, thick}
        \legend{
            Backend -- w/ central backend,
            UAS -- w/ central backend,
            World -- active UAS,
            Backend -- w/ central backend,
            UAS -- w/ central backend,
            World -- active UAS
            }
        \end{axis}
    \end{tikzpicture}
    \caption{
        Number of known UAS for one UAS and the backend in the scenario w/ central backend using 81 ground stations.
    }
    \label{fig:vehcountplot}    
\end{figure}
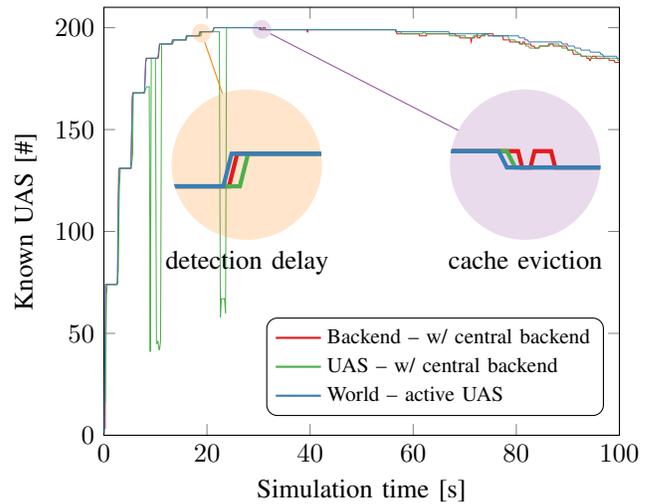

\begin{table}
	\centering
	\caption{Comparison of the \ac{EAR}, payload length, and channel load for the five scenarios. For the scenario w/ central backend, results are separated between \ac{UAS} and \acp{GS}.}
	\label{tab:communication_results}
	\setlength{\tabcolsep}{2.5pt}%
\renewcommand{\arraystretch}{1.2}  %
\centering
\begin{tabularx}{\linewidth}{l r r r r r r r r r}
	\toprule
	\textbf{Scenario} & \multicolumn{2}{c}{\textbf{EAR} $[\%]$} & \multicolumn{3}{c}{\textbf{Payload} $[Byte]$} & \multicolumn{3}{c}{\textbf{Channel load} $[\%]$} \\
	& \multicolumn{1}{c}{avg} & \multicolumn{1}{c}{max} & \multicolumn{1}{c}{min} & \multicolumn{1}{c}{avg} & \multicolumn{1}{c}{max} & \multicolumn{1}{c}{min} & \multicolumn{1}{c}{avg} & \multicolumn{1}{c}{max} \\
	\midrule

	\textbf{Local perception} &
	3.46 & %
	16.67 & %
	0 & %
	0 & %
	0 & %
	0.02 & %
	0.04 & %
	0.06 %
	\\

	\textbf{CA} &
	13.39 & %
	25.26 & %
	41 & %
	103 & %
	241 & %
	2.35 & %
	4.20 & %
	5.77 %
	\\

	\textbf{CP} &
	26.13 & %
	55.14 & %
	46 & %
	241 & %
	945 & %
	6.86 & %
	15.34 & %
	22.80 %
	\\

	\textbf{CA \& CP} &
	26.86 & %
	58.03 & %
	41 & %
	193 & %
	945 & %
	8.96 & %
	16.58 & %
	22.51 %
	\\

	\multicolumn{10}{l}{\textbf{w/ central backend and 81 \acfp{GS}}} \\
	\textbf{\hspace{0.2cm}\rotatebox[origin=c]{180}{\contour{black}{$\Lsh$}} UAS} &
	66.24 & %
	100.51 & %
	41 & %
	192 & %
	945 & %
	11.40 & %
	19.49 & %
	25.73 %
	\\

	\textbf{\hspace{0.2cm}\rotatebox[origin=c]{180}{\contour{black}{$\Lsh$}} \acp{GS}} &
	99.56 & %
	100.51 & %
	1661 & %
	5538 & %
	5837 & %
	8.64 & %
	16.90 & %
	21.80 %
	\\

	\bottomrule
\end{tabularx}

\end{table}

Fig.~\ref{fig:vehcountplot} further highlights the detection delay and caching, showing the number of known \acp{UAS} w/ central backend.
\acp{UAS} known by a \ac{UAS}, and the backend are compared to the active \acp{UAS} in the simulation world.
The detection follows the number of active \acp{UAS}.
The 200 \acp{UAS} spawn in the first \SI{20}{\second} and start despawning in the last \SI{20}{\second}.

The delay between spawning and detection of a \ac{UAS} is determined by the \ac{CP} message frequency of \SIrange{1}{10}{\hertz}, which reports objects that arrive.
The backend aggregates the objects and includes them in the next central \ac{CP} message, again at \SIrange{1}{10}{\hertz}. 
Thus, the worst case for the detection time with central \ac{CP} is between \SIrange{0.2}{2}{\second} plus a transmission delay.

The previously mentioned timeout causes a maximum \ac{EAR} higher than 100\% shown in Table~\ref{tab:communication_results} as despawning \ac{UAS} are still cached.
Cache eviction occurs when a \ac{UAS} is not detected by sensing, \ac{CA}, or \ac{CP} and is removed from the cache after \SI{1.1}{\second}. 
This timeout reflects the freshness requirements for object information and depends on the minimum frequency for \ac{CP} messages of \SI{1}{\hertz} (\cf Table~\ref{tab:vehicle_information}) plus a 10\% buffer for transmission delays, etc.
The central backend increases this effect, extending the live time of detected objects by re-advertising. 
\ac{UAS} have no impact on this as they only share objects detected with their own sensors.
Further investigation can help to optimize the caching strategy.

\section{Conclusion and Outlook}%
\label{sec:conclusion_and_outlook}

In this work, we explored the \ac{CP} for \ac{UAS} in \ac{UAM}. 
Based on existing standards from the aviation, automotive, and drone domains, we identified a \ac{CP} data space and the desired frequency of information exchange. 
We proposed a hybrid approach for \ac{CP} using local broadcast combined with a central \ac{CP} service. 
\acp{GS} forward local \ac{CP} messages that are collected and aggregated at a centralized service to be then shared with airspace users again. 

In a case study, we evaluated the environmental awareness of the proposed \ac{CP} service through simulations, comparing it to distributed approaches. 
We analyzed the impact of the number of \acp{GS} and the frequency of \ac{CP} broadcasts on the \ac{EAR} and the communication channel load.
Our results show that with increased density of \acp{GS} the channel load increases, but the \ac{EAR} also improves until the messages are reduced due to channel congestion.
Continuing with a spacing of \SI{500}{\meter} between GS, the centralized approach significantly improved the average \ac{EAR} to 99\% in the backend service and 66\% at \ac{UAS} from just 3.5\% without communication and 27\% with distributed \ac{CP}.
In turn, the channel load is also increased from an average of 16.5\% with distributed \ac{CP} to 19.5\% with the centralized \ac{CP} service. 
The increased environmental awareness can help with advanced conflict detection and resolution, enhancing the overall safety of \ac{UAM} operations~\cite{rhjgk-cdtjr-24}.

Our evaluation raised central research questions about the optimal \ac{CP} broadcast range, frequency, and ideal \ac{GS} number and placement.
Future work shall investigate additional \ac{CP} metrics such as the age of information, object redundancy, and consider the zone of interest for a \ac{UAS}. 
Further improvements can be achieved through more realistic simulation scenarios and environments, \eg by providing realistic \ac{UAM} traffic models.
Our findings can aid in developing solutions for cooperative conflict detection and resolution in U-space airspace, aiding in identifying challenging objects such as animals or uncooperative entities~\cite{rhjgk-cdtjr-24}. 
Integrating knowledge about these objects into operations can enhance the overall safety of all airspace users by deliberately avoiding them.

\bibliographystyle{IEEEtran}

\end{document}